\documentclass[11pt]{elsart}
\usepackage{amsfonts}
\usepackage{slashbox}
\usepackage{graphicx}
\usepackage{amssymb,amsmath}
\usepackage{mathrsfs} 

\begin{document}

\begin{frontmatter}
\title{On amending the Maskin's sufficiency theorem by using complex numbers}
\author{Haoyang Wu\corauthref{cor}}
\corauth[cor]{Wan-Dou-Miao Research Lab, Suite 1002, 790 WuYi Road,
Shanghai, 200051, China.} \ead{hywch@mail.xjtu.edu.cn} \ead{Tel:
86-18621753457}

\begin{abstract}
The Maskin's theorem is a fundamental work in the theory of
mechanism design. In this paper, we will propose a self-enforcing
agreement by which agents may break through the Maskin's sufficiency
theorem if the designer uses the Maskin's mechanism, i.e,. a social
choice rule which satisfies monotonicity and no-veto may be not Nash
implementable. The agreement is based on an algorithm with complex
numbers. It is justified when the designer communicates with the
agents through some channels (e.g., Internet). Since the designer
cannot prevent the agents from signing such self-enforcing
agreement, the Maskin's sufficiency theorem is amended.
\end{abstract}
\end{frontmatter}

\section{Introduction}
Nash implementation is the cornerstone of the mechanism design
theory. The Maskin's theorem provides an almost complete
characterization of social choice rules (SCRs) that are Nash
implementable: When the number of agents are at least three, the
sufficient conditions for Nash implementation are monotonicity and
no-veto, and the necessary condition is monotonicity
\cite{Maskin1999}. Note that an SCR is specified by a designer, a
desired outcome from the designer's perspective may not be desirable
for the agents. However, when the number of agents are at least
three, by the Maskin's theorem the designer can always implement an
SCR which satisfies monotonicity and no-veto in Nash equilibrium
even if all agents dislike it (See Table 1 in Section 3.1).

With the development of network economics, it is more and more
common that the designer communicates with agents through some
channel (e.g., Internet). For this case, we will show that the
agents may find a way to break through the restriction of the
Maskin's sufficiency theorem. Suppose that the agents face a bad SCR
that satisfies monotonicity and no-veto, and the designer claims the
traditional Maskin's mechanism. We will propose a self-enforcing
agreement by which agents can make the SCR not Nash implementable if
an additional condition is satisfied.

The rest of the paper is organized as follows: Section 2 recalls
preliminaries of the mechanism design theory \cite{Serrano2004};
Section 3 is the main part of this paper, where we will propose a
self-enforcing agreement using complex numbers to amend the Maskin's
sufficiency theorem. Section 4 draws conclusions.

\section{Preliminaries}
Let $N=\{1,\cdots,n\}$ be a finite set of \emph{agents} with $n\geq
2$, $A=\{a_{1},\cdots,a_{k}\}$ be a finite set of social
\emph{outcomes}. Suppose each agent $j$ privately observes a
parameter $t_{j}$ that determines his preferences over the outcomes
in $A$. We refer to $t_{j}$ as agent $j$'s \emph{type}. The set of
possible types for agent $j$ is denoted as $T_{j}$. We refer to a
profile of types $t=(t_{1},\cdots,t_{n})$ as a \emph{state}. Let
$\mathcal {T}=\prod_{j\in N}T_{j}$ be the set of states. At state
$t\in\mathcal {T}$, each agent $j\in N$ is assumed to have a
complete and transitive \emph{preference relation} $\succeq_{j}^{t}$
over the set $A$. We denote by
$\succeq^{t}=(\succeq_{1}^{t},\cdots,\succeq_{n}^{t})$ the profile
of preferences in state $t$, and denote by $\succ_{j}^{t}$ the
strict preference part of $\succeq_{j}^{t}$. Fix a state $t$, we
refer to the collection $E=<N,A,(\succeq_{j}^{t})_{j\in N}>$ as an
\emph{environment}. Let $\varepsilon$ be the class of possible
environments. A \emph{social choice rule} (SCR) $F$ is a mapping
$F:\varepsilon\rightarrow 2^{A}\backslash\{\emptyset\}$. A
\emph{mechanism} $\Gamma=((M_{j})_{j\in N},g)$ describes a message
or strategy set $M_{j}$ for agent $j$, and an outcome function
$g:\prod_{j\in N}M_{j}\rightarrow A$. $M_{j}$ is unlimited except
that if a mechanism is direct, \emph{i.e.}, $M_{j}=T_{j}$.

An SCR $F$ satisfies \emph{no-veto} if, whenever $a\succeq_{j}^{t}b$
for all $b\in A$ and for every agent $j$ but perhaps one $k$, then
$a\in F(E)$. An SCR $F$ is \emph{monotonic} if for every pair of
environments $E$ and $E'$, and for every $a\in F(E)$, whenever
$a\succeq_{j}^{t}b$ implies that $a\succeq_{j}^{t'}b$, there holds
$a\in F(E')$. We assume that there is \emph{complete information}
among the agents, \emph{i.e.}, the true state $t$ is common
knowledge among them. Given a mechanism $\Gamma=((M_{j})_{j\in
N},g)$ played in state $t$, a \emph{Nash equilibrium} of $\Gamma$ in
state $t$ is a strategy profile $m^{*}$ such that: $\forall j\in N,
g(m^{*}(t))\succeq_{j}^{t}g(m_{j},m_{-j}^{*}(t)), \forall m_{j}\in
M_{j}$. Let $\mathcal {N}(\Gamma,t)$ denote the set of Nash
equilibria of the game induced by $\Gamma$ in state $t$, and
$g(\mathcal {N}(\Gamma,t))$ denote the corresponding set of Nash
equilibrium outcomes. An SCR $F$ is \emph{Nash implementable} if
there exists a mechanism $\Gamma=((M_{j})_{j\in N},g)$ such that for
every $t\in \mathcal {T}$, $g(\mathcal {N}(\Gamma,t))=F(t)$.

Maskin \cite{Maskin1999} provided an almost complete
characterization of SCRs that were Nash implementable. The main
results of Ref. \cite{Maskin1999} are two theorems: 1)
(\emph{Necessity}) If an SCR is Nash implementable, then it is
monotonic. 2) (\emph{Sufficiency}) Let $n\geq3$, if an SCR is
monotonic and satisfies no-veto, then it is Nash implementable. In
order to facilitate the following investigation, we briefly recall
the Maskin's mechanism published in Ref. \cite{Serrano2004} as
follows:

Consider the following mechanism $\Gamma=((M_{j})_{j\in N},g)$,
where agent $j$'s message set is $M_{j}=A\times \mathcal {T} \times
\mathbb{Z}_{+}$, where $\mathbb{Z}_{+}$ is the set of non-negative
integers. A typical message sent by agent $j$ is described as
$m_{j}=(a_{j},t_{j},z_{j})$. The outcome function $g$ is defined in
the following three rules: (1) If for every agent $j\in N$,
$m_{j}=(a,t,0)$ and $a\in F(t)$, then $g(m)=a$. (2) If $(n-1)$
agents $j\neq k$ send $m_{j}=(a,t,0)$ and $a\in F(t)$, but agent $k$
sends $m_{k}=(a_{k},t_{k},z_{k})\neq(a,t,0)$, then $g(m)=a$ if
$a_{k}\succ_{k}^{t}a$, and $g(m)=a_{k}$ otherwise. (3) In all other
cases, $g(m)=a'$, where $a'$ is the outcome chosen by the agent with
the lowest index among those who announce the highest integer.

\section{Amending the Maskin's sufficiency theorem}
This section is the main part of this paper. In the beginning, we
will show a bad SCR which satisfies monotonicity and no-veto. It is
Nash implementable although all agents dislike it. Then, we will
define some matrices and propose a self-enforcing agreement using
complex numbers, by which the agents can amend the Maskin's
sufficiency theorem and make the bad SCR not Nash implementable.

\subsection{A bad SCR}
\emph{Table 1: A bad SCR that satisfies monotonicity and no-veto.}\\
\begin{tabular}{cccccc}
 \multicolumn{3}{c}{State $t^{1}$}&\multicolumn{3}{c}{State $t^{2}$}\\
 $Apple$&$Lily$ &$Cindy$ &$Apple$&$Lily$ &$Cindy$\\ \hline
 $a^{3}$&$a^{2}$&$a^{1}$ &$a^{4}$&$a^{3}$&$a^{1}$ \\
 $a^{1}$&$a^{1}$&$a^{3}$ &$a^{1}$&$a^{1}$&$a^{2}$ \\
 $a^{2}$&$a^{4}$&$a^{2}$ &$a^{2}$&$a^{2}$&$a^{3}$ \\
 $a^{4}$&$a^{3}$&$a^{4}$ &$a^{3}$&$a^{4}$&$a^{4}$ \\\hline
 \multicolumn{3}{c}{$F(t^{1})=\{a^{1}\}$}&\multicolumn{3}{c}{$F(t^{2})=\{a^{2}\}$}\\\hline
\end{tabular}

Let $N=\{Apple, Lily, Cindy\}$, $\mathcal {T}=\{t^{1},t^{2}\}$,
$A=\{a^{1},a^{2},a^{3},a^{4}\}$. In each state $t\in\mathcal {T}$,
the preference relations $(\succeq^{t}_{j})_{j\in N}$ over the
outcome set $A$ and the corresponding SCR $F$ are given in Table 1.
The SCR $F$ is bad from the agents' perspectives because in state
$t^{2}$, all agents unanimously prefer a Pareto-efficient outcome
$a^{1}\in F(t^{1})$: for each agent $j$,
$a^{1}\succ^{t^{2}}_{j}a^{2}\in F(t^{2})$.

At first sight, in state $t^{2}$, $(a^{1},t^{1},0)$ should be a
unanimous $m_{j}$ for each agent $j$, because by doing so $a^{1}$
would be generated by rule 1. However, $Apple$ has an incentive to
unilaterally deviate from $(a^{1},t^{1},0)$ to $(a^{4},*,*)$ in
order to trigger rule 2, since $a^{1}\succ^{t^{1}}_{Apple}a^{4}$,
$a^{4}\succ^{t^{2}}_{Apple}a^{1}$; $Lily$ also has an incentive to
unilaterally deviate from $(a^{1},t^{1},0)$ to $(a^{3},*,*)$, since
$a^{1}\succ^{t^{1}}_{Lily}a^{3}$, $a^{3}\succ^{t^{2}}_{Lily}a^{1}$.

Note that either $Apple$ or $Lily$ can certainly obtain her expected
outcome only if just one of them deviates from $(a^{1},t^{1},0)$ (If
this case happened, rule 2 would be triggered). But this condition
is unreasonable, because all agents are rational, nobody is willing
to give up and let the others benefit. Therefore, both $Apple$ and
$Lily$ will deviate from $(a^{1},t^{1},0)$. As a result, rule 3 will
be triggered. Since $Apple$ and $Lily$ both have a chance to win the
integer game, the final winner is uncertain. Consequently, the final
outcome is uncertain between $a^{3}$ and $a^{4}$.

To sum up, although every agent prefers $a^{1}$ to $a^{2}$ in state
$t^{2}$, $a^{1}$ cannot be yielded in Nash equilibrium. Indeed, the
Maskin's mechanism makes the Pareto-inefficient outcome $a^{2}$ be
implemented in Nash equilibrium in state $t^{2}$.

\emph{Can the agents find a way to let the Pareto-efficient outcome
$a^{1}$ be Nash implemented in state $t^{2}$ when the designer uses
the Maskin's mechanism?} Interestingly, we will show that the answer
may be ``yes''. To do so, a new weapon - the complex number - will
be used. Although it has been well-known for hundreds of years, it
has never been used in the theory of mechanism design. In what
follows, first we will define some matrices with complex numbers,
then we will propose a self-enforcing agreement to help agents break
through the Maskin's sufficiency theorem.

\subsection{Definitions}
\textbf{Definition 1}: Let $\hat{I}, \hat{\sigma}$ be two $2\times
2$ matrices, and $\overrightarrow{C}, \overrightarrow{D}$ be two
basis vectors:
\begin{equation} \quad \hat{I}\equiv\begin{bmatrix}
  1 & 0\\
  0 & 1
\end{bmatrix},\quad \hat{\sigma}\equiv\begin{bmatrix}
  0 & 1\\
  1 & 0
\end{bmatrix}, \overrightarrow{C}\equiv\begin{bmatrix}
  1\\
  0
\end{bmatrix},\quad \overrightarrow{D}\equiv\begin{bmatrix}
  0\\
  1
\end{bmatrix}.
\end{equation}
Hence, $\hat{I}\overrightarrow{C}=\overrightarrow{C}$,
$\hat{I}\overrightarrow{D}=\overrightarrow{D}$;
$\hat{\sigma}\overrightarrow{C}=\overrightarrow{D}$,
$\hat{\sigma}\overrightarrow{D}=\overrightarrow{C}$.

\textbf{Definition 2}: For $n\geq 3$ agents, suppose each agent
$j\in N$ possess a basis vector. $\overrightarrow{\psi}_{0}$ is
defined as the tensor product of $n$ basis vectors
$\overrightarrow{C}$:
\begin{equation}
\overrightarrow{\psi}_{0}\equiv\overrightarrow{C}^{\otimes
n}\equiv\underbrace{\overrightarrow{C}\otimes\cdots \otimes
\overrightarrow{C}}\limits_{n}\equiv\begin{bmatrix}
  1\\
  0\\
  \cdots\\
  0
\end{bmatrix}_{2^{n}\times 1}
\end{equation}
$\overrightarrow{C}^{\otimes n}$ contains $n$ basis vectors
$\overrightarrow{C}$ and $2^{n}$ elements.
$\overrightarrow{C}^{\otimes n}$ is also denoted as
$\overrightarrow{C\cdots CC}^{n}$. Similarly,
\begin{equation}
\overrightarrow{C\cdots
CD}^{n}\equiv\underbrace{\overrightarrow{C}\otimes\cdots \otimes
\overrightarrow{C}}\limits_{n-1}\otimes
\overrightarrow{D}=\begin{bmatrix}
  0\\
  1\\
  \cdots\\
  0
\end{bmatrix}_{2^{n}\times 1}
\end{equation}
Obviously, there are $2^{n}$ possible vectors
$\{\overrightarrow{C\cdots CC}^{n},$ $
\cdots,\overrightarrow{D\cdots DD}^{n}\}$.

\textbf{Definition 3}: $\hat{J}\equiv
\frac{1}{\sqrt{2}}(\hat{I}^{\otimes n}+i\hat{\sigma}^{\otimes  n})$,
\emph{i.e.},
\begin{equation}
\hat{J}\equiv\frac{1}{\sqrt{2}}\begin{bmatrix}
  1 &  &  &  &  &  & i\\
   & \cdots  & &  & & \cdots  & \\
   &  &  & 1 & i &  & \\
   &  &  & i & 1 &  & \\
   & \cdots  & &  &  & \cdots & \\
  i &  &  &  &  &  & 1
\end{bmatrix}_{2^{n}\times2^{n}},
\hat{J}^{+}\equiv\frac{1}{\sqrt{2}}\begin{bmatrix}
  1 &  &  &  &  &  & -i\\
   & \cdots  & &  & & \cdots  & \\
   &  &  & 1 & -i &  & \\
   &  &  & -i & 1 &  & \\
   & \cdots  & &  &  & \cdots & \\
  -i &  &  &  &  &  & 1
\end{bmatrix}_{2^{n}\times2^{n}}
\end{equation}
where the symbol $i$ denotes an imaginary number, and $\hat{J}^{+}$
is the conjugate transpose of $\hat{J}$. In what follows, we will
not explicitly claim whether $i$ is an imaginary number or an index.
It is easy for the reader to know its exact meaning from the
context.

\textbf{Definition 4}:
\begin{equation}
\overrightarrow{\psi}_{1}\equiv
\hat{J}\overrightarrow{\psi}_{0}=\frac{1}{\sqrt{2}}\begin{bmatrix}
  1\\
  0\\
  \cdots\\
  0\\
  i
\end{bmatrix}_{2^{n}\times 1}
\end{equation}
\textbf{Definition 5}: For $\theta\in[0, \pi]$, $\phi\in [0,
\pi/2]$,
\begin{equation}
  \hat{\omega}(\theta, \phi)\equiv\begin{bmatrix}
  e^{i\phi}\cos(\theta/2) & i\sin(\theta/2)\\
  i\sin(\theta/2) & e^{-i\phi}\cos(\theta/2)
\end{bmatrix}.
\end{equation}
$\hat{\Omega}\equiv\{\hat{\omega}(\theta,\phi):\theta\in[0,\pi],\phi\in[0,\pi/2]\}$.
Hence, $\hat{I}=\hat{\omega}(0, 0)$,
$\hat{\sigma}=-i\hat{\omega}(\pi, 0)$.

 \textbf{Definition 6}: For $j=1, \cdots, n$, $\theta_{j}\in[0, \pi]$, $\phi_{j}\in [0,
\pi/2]$, let  $\hat{\omega}_{j}=\hat{\omega}(\theta_{j}, \phi_{j})$,
\begin{equation}
\overrightarrow{\psi}_{2}\equiv[\hat{\omega}_{1}\otimes\cdots\otimes\hat{\omega}_{n}]\overrightarrow{\psi}_{1}.
\end{equation}
The dimension of
$\hat{\omega}_{1}\otimes\cdots\otimes\hat{\omega}_{n}$ is
$2^{n}\times 2^{n}$. Since only two elements in
$\overrightarrow{\psi}_{1}$ are non-zero, it is not necessary to
calculate the whole $2^{n}\times2^{n}$ matrix to yield
$\overrightarrow{\psi}_{2}$. Indeed, we only need to calculate the
leftmost and rightmost column of
$\hat{\omega}_{1}\otimes\cdots\otimes\hat{\omega}_{n}$ to derive
$\overrightarrow{\psi}_{2}$.

\textbf{Definition 7}: $\overrightarrow{\psi}_{3}\equiv
\hat{J}^{+}\overrightarrow{\psi}_{2}$.

Suppose $\overrightarrow{\psi}_{3}=[\eta_{1}, \cdots,
\eta_{2^{n}}]^{T}$, let $\Delta=[|\eta_{1}|^{2}, \cdots,
|\eta_{2^{n}}|^{2}]$. It can be easily checked that $\hat{J}$,
$\hat{\omega}_{j}$ ($j=1,\cdots, n$) and $\hat{J}^{+}$ are all
unitary matrices. Hence, $|\overrightarrow{\psi}_{3}|^{2}=1$. Thus,
$\Delta$ can be viewed as a probability distribution, each element
of which represents the probability that we randomly choose a vector
from the set of all $2^{n}$ possible vectors
$\{\overrightarrow{C\cdots CC}^{n},$ $
\cdots,\overrightarrow{D\cdots DD}^{n}\}$.

\textbf{Definition 8}: Condition $\lambda$ contains
five parts. The first three parts are defined as follows:\\
$\lambda_{1}$: Given an SCR $F$, there exist two states $\hat{t}$,
$\bar{t}\in \mathcal {T}$, $\hat{t}\neq \bar{t}$ such that
$\hat{a}\succeq^{\bar{t}}_{j}\bar{a}$ (for each $j\in N$,
$\hat{a}\in F(\hat{t})$, $\bar{a}\in F(\bar{t})$) with strict
relation for some agent; and the number of agents that encounter a
preference change around $\hat{a}$ in going from state $\hat{t}$ to
$\bar{t}$ is at least two. Denote by $l$ the number of these agents.
Without loss of generality, let these $l$ agents be the last $l$
agents among $n$ agents, \emph{i.e.}, agent $(n-l+1),\cdots, n$.\\
$\lambda_{2}$: Consider the state $\bar{t}$ specified in condition
$\lambda_{1}$, if there exists another $\hat{t}'\in \mathcal {T}$,
$\hat{t}'\neq\hat{t}$ that satisfies $\lambda_{1}$, then
$\hat{a}\succeq^{\bar{t}}_{j}\hat{a}'$ (for each $j\in N$,
$\hat{a}\in F(\hat{t})$, $\hat{a}'\in
F(\hat{t}')$) with strict relation for some agent.\\
$\lambda_{3}$: Consider the outcome $\hat{a}$ specified in condition
$\lambda_{1}$, for any state $t\in\mathcal{T}$, $\hat{a}$ is top
ranked for each agent $j$ among the first $(n-l)$ agents.

\subsection{An agreement that uses complex numbers}
As we have seen, the Maskin's mechanism is an abstract mechanism.
People seldom consider the manner in which the designer actually
receives messages from agents. Roughly speaking, there are two
manners: direct and indirect manner. In a direct manner, agents
report their messages to the designer directly (\emph{e.g.}, by
hand, or face to face etc), thereby the designer can be sure that
any message is submitted by an agent himself, not by any other
device. In an indirect manner, the agents report messages to the
designer through some channels (\emph{e.g.}, Internet). Therefore,
when the designer receives a message from a channel, he cannot know
what has happened on the other side of the channel. Put differently,
the designer cannot discriminate whether the message is submitted by
an agent himself, or generated by some other device. In what
follows, we assume the designer receives messages from the agents in
an indirect manner.

\textbf{Definition 9}: Suppose conditions $\lambda_{1}$,
$\lambda_{2}$ and $\lambda_{3}$ are satisfied, and the designer uses
the Maskin's mechanism. An agreement \emph{ComplexMessage} is
constructed by the agents (see Fig. 1). It is constructed after the
designer claims the outcome function $g$, and before the designer
receives messages $m=(m_{1}, \cdots, m_{n})$ from agents indirectly.
The algorithm \emph{MessageComputing} is given in Definition 10.

\begin{figure}[!t]
\centering
\includegraphics[height=2.8in,clip,keepaspectratio]{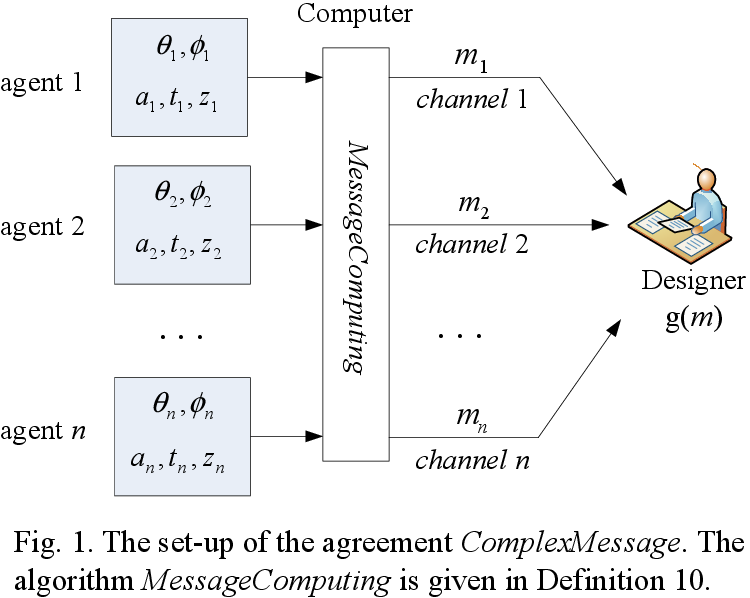}
\end{figure}

\textbf{Definition 10}: The algorithm \emph{MessageComputing} is defined as follows:\\
\textbf{Input}: $(\theta_{j}, \phi_{j}, a_{j},t_{j},z_{j})\in
[0,\pi/2]\times [0,\pi]\times A\times \mathcal {T} \times
\mathbb{Z}_{+}$,
$j=1,\cdots,n$.\\
\textbf{Output}:  $m_{j}\in A\times \mathcal {T} \times
\mathbb{Z}_{+}$, $j=1,\cdots,n$.\\
1: Reading $(\theta_{j},
\phi_{j})$ from each agent $j\in N$ (See Fig. 2(a)).\\
2: Computing the leftmost and rightmost columns of
$\hat{\omega}_{1}\otimes\cdots\otimes\hat{\omega}_{n}$ (See Fig. 2(b)).\\
3: Computing
$\overrightarrow{\psi}_{2}=[\hat{\omega}_{1}\otimes\cdots\otimes\hat{\omega}_{n}]\overrightarrow{\psi}_{1}$,
$\overrightarrow{\psi}_{3}=\hat{J}^{+}\overrightarrow{\psi}_{2}$,
and the probability distribution
$\Delta$ (See Fig. 2(c)).\\
4: Randomly choosing a vector from the set of all $2^{n}$ possible
vectors $\{\overrightarrow{C\cdots CC}^{n},$ $
\cdots,\overrightarrow{D\cdots DD}^{n}\}$ according to the
probability distribution $\Delta$.\\
5: For each agent $j\in N$, let $m_{j}=(\hat{a}, \hat{t}, 0)$ (or
$m_{j}=(a_{j},t_{j},z_{j})$) if the $j$-th basis vector of the
chosen vector
is $\overrightarrow{C}$ (or $\overrightarrow{D}$) (See Fig. 2(d)).\\
6: Sending $m=(m_{1}, \cdots, m_{n})$ to the designer through
channels $1, \cdots, n$.

When \emph{ComplexMessage} has been constructed, it can be seen from
Fig. 1 that all agents has transferred their channels to the
computer. After then, each agent $j\in N$ can leave his channel to
the computer, or
take back his channel and send his message to the designer directly:\\
1) Whenever any agent takes back his channel, every other agent will
detect this deviation and take back their channels too. Thereby,
all agents will send their messages to the designer directly. \\
2) When all agents leave their channels to the computer, the
algorithm \emph{MessageComputing} works, i.e., calculates $m=(m_{1},
\cdots, m_{n})$ and sends it to the designer.\\
Put differently, after \emph{ComplexMessage} is
constructed, each agent $j\in N$ independently faces two options:\\
$\bullet$ $S(j,0)$: leaving his channel to the computer, and
submitting
$(\theta_{j}, \phi_{j}, a_{j}, t_{j}, z_{j})$ to the algorithm \emph{MessageComputing}.\\
$\bullet$ $S(j,1)$: taking back his channel, and submitting $(a_{j},
t_{j}, z_{j})$ to the designer directly.

To sum up, suppose the agents sign the agreement
\emph{ComplexMessage} after the designer claims the outcome function
$g$, the timing steps of the mechanism are updated as follows:\\
Time 1: The designer claims the outcome function $g$ to all agents;\\
Time 2: The agents sign the agreement \emph{ComplexMessage};\\
Time 3: Each agent $j\in N$ chooses an option between $S(j,0)$ and
$S(j,1)$.\\
Time 4: The designer receives $m=(m_{1}, \cdots, m_{n})$ from $n$ channels; \\
Time 5: The designer computes the outcome $g(m)$.

\textbf{Remark 1:} Although the time and space complexity of
\emph{MessageComputing} are exponential, \emph{i.e.}, $O(2^{n})$, it
works well when the number of agents is not large. For example, the
runtime of \emph{MessageComputing} is about 0.5s for 15 agents, and
about 12s for 20 agents (MATLAB 7.1, CPU: Intel (R) 2GHz, RAM: 3GB).

\textbf{Remark 2:} The problem of Nash implementation requires
complete information among all agents. In the last paragraph of Page
392 \cite{Serrano2004}, Serrano wrote: ``\emph{We assume that there
is complete information among the agents... This assumption is
especially justified when the implementation problem concerns a
small number of agents that hold good information about one
another}''. Hence, the fact that \emph{MessageComputing} is suitable
for small-scale cases (\emph{e.g.}, less than 20 agents) is
acceptable for Nash implementation.

\textbf{Definition 11}: Consider the state $\bar{t}$ specified in
condition $\lambda_{1}$. Suppose $\lambda_{1}$ and $\lambda_{2}$ are
satisfied, and $m=(m_{1}, \cdots, m_{m})$ is computed by
\emph{MessageComputing}. $\$_{C\cdots CC}$, $\$_{C\cdots CD}$,
$\$_{D\cdots DC}$ and $\$_{D\cdots DD}$ are defined as the payoffs
to the $n$-th agent in state $\bar{t}$ when the chosen vector in
Step 4 of \emph{MessageComputing} is $\overrightarrow{C\cdots
CC}^{n}$, $\overrightarrow{C\cdots CD}^{n}$,
$\overrightarrow{D\cdots DC}^{n}$ or $\overrightarrow{D\cdots
DD}^{n}$ respectively.

\textbf{Definition 12}: Suppose conditions $\lambda_{1}$,
$\lambda_{2}$ and $\lambda_{3}$ are satisfied.  When the true state
is $\bar{t}$, consider each message $m_{j}=(a_{j}, t_{j}, z_{j})$,
where $a_{j}$ is top-ranked for each agent $j$. The rest two parts
of condition
$\lambda$ are defined as:\\
$\lambda_{4}$: $\$_{C\cdots CC}>\$_{D\cdots DD}$.\\
$\lambda_{5}$: $\$_{C\cdots CC}>\$_{C\cdots
CD}\cos^{2}(\pi/l)+\$_{D\cdots DC}\sin^{2}(\pi/l)$.\\

\subsection{Main result}
\textbf{Proposition 1}: For $n\geq 3$, suppose the agents send
messages to the designer indirectly. Consider an SCR $F$ that
satisfies monotonicity and no-veto. Suppose the designer uses the
Maskin's mechanism $\Gamma$ and condition $\lambda$ is satisfied,
then in state $\bar{t}$ the agents can sign the agreement
\emph{ComplexMessage} to make the Pareto-inefficient outcome
$F(\bar{t})$ not be yielded in Nash equilibrium.

\textbf{Proof}: Since $\lambda_{1}$ and $\lambda_{2}$ are satisfied,
then there exist two states $\hat{t}$, $\bar{t}\in \mathcal {T}$,
$\hat{t}\neq \bar{t}$ such that
$\hat{a}\succeq^{\bar{t}}_{j}\bar{a}$ (for each $j\in N$,
$\hat{a}\in F(\hat{t})$, $\bar{a}\in F(\bar{t})$) with strict
relation for some agent; and the number of agents that encounter a
preference change around $\hat{a}$ in going from state $\hat{t}$ to
$\bar{t}$ is at least two. Suppose the true state is $\bar{t}$, now
let us check whether the agents can make the Pareto-inefficient
outcome $\bar{a}$ not be implemented in Nash equilibrium by
constructing \emph{ComplexMessage}.

Note that after the agents construct \emph{ComplexMessage}, in Time
4 the designer cannot discriminate whether the received messages
$(m_{1},\cdots,m_{n})$ are submitted by agents themselves or sent by
\emph{MessageComputing}. However, from the viewpoints of agents, the
situation is different from the traditional Maskin's mechanism.
After constructing \emph{ComplexMessage}, there are two
possible cases in Time 3:\\
1) Suppose every agent $j$ chooses $S(j,0)$, then the algorithm
\emph{MessageComputing} works. Consider the following strategy
profile chosen by the agents: each agent $j=1, \cdots, (n-l)$
submits $(\theta_{j}, \phi_{j})=(0, 0)$; each agent $j=(n-l+1),
\cdots, n$ submits $(\theta_{j}, \phi_{j})=(0,\pi/l)$. Since
condition $\lambda$ is satisfied, according to Lemma 1 (see
Appendix), this strategy profile is a Nash equilibrium of $\Gamma$
in state $\bar{t}$. As a result, in Step 4 of
\emph{MessageComputing}, the chosen vector will be
$\overrightarrow{C\cdots CC}$; in Step 5 of \emph{MessageComputing},
$m_{j}=(\hat{a}, \hat{t}, 0)$ for each $j\in N$. In Time 5,
$g(m)=\hat{a}\notin F(\bar{t})$. Each agent $j$'s payoff is
$\$_{C\cdots CC}$.\\
2) Suppose some agent $j\in N$ chooses $S(j,1)$, \emph{i.e.}, takes
back his channel and reports $m_{j}$ to the designer directly. Then
all of the rest agents will observe this deviation, thereby take
back their channels and submit messages to the designer directly. In
Time 5, the final outcome implemented in Nash equilibrium will be
$F(\bar{t})$, and each agent $j$'s payoff is $\$_{D\cdots DD}$.

Since condition $\lambda_{4}$ is satisfied, it is not profitable for
any agent $j$ to unilaterally take back his channel and send a
message to the designer directly. According to Telser
\cite{Telser1980}, \emph{ComplexMessage} is a self-enforcing
agreement among the agents. Put differently, although the agents
collaborate to construct \emph{ComplexMessage} in Time 2, they do
not require a third-party to enforce it after then.

To sum up, in state $\bar{t}$, the agents can sign a self-enforcing
agreement \emph{ComplexMessage} to make the Pareto-inefficient
outcome $F(\bar{t})$ not be implemented in Nash equilibrium.
$\square$

\section{Conclusions}
In this paper, we propose a self-enforcing agreement to help agents
avoid the Pareto-inefficient outcome when they face a bad social
choice rule. When the designer uses the Maskin's mechanism and
receives messages from the agents indirectly (e.g., Internet), the
designer cannot restrict the agents from signing such agreement. It
should be noted that the introduction of complex numbers plays an
important role in this paper. To the best of our knowledge, there is
no similar work before. Since the Maskin's mechanism has been widely
applied to many disciplines, there are many works to do in the
future to generalize the self-enforcing agreement further.

%


\newpage
\section*{Appendix}
\textbf{Lemma 1}: Suppose the algorithm \emph{MessageComputing}
works. If condition $\lambda$ is satisfied, consider the following strategy:\\
1) Each agent $j=1, \cdots, (n-l)$ submits $(\theta_{j},
\phi_{j})=(0, 0)$; \\
2) Each agent $j=(n-l+1), \cdots, (n-1)$ submits $(\theta_{j},
\phi_{j})=(0,\pi/l)$;\\
then the optimal value of $(\theta,\phi)$ for the $n$-th agent is
$(0,\pi/l)$.

\textbf{Proof}: Since condition $\lambda_{1}$ is satisfied, then
$l\geq 2$. Let
\begin{equation*}
  \hat{C}_{l}\equiv\hat{\omega}(0,\pi/l)=\begin{bmatrix}
  e^{i\frac{\pi}{l}} & 0 \\
  0 & e^{-i\frac{\pi}{l}}
\end{bmatrix}_{2\times 2},\quad\mbox{ thus, }
  \hat{C}_{l}\otimes\hat{C}_{l}=\begin{bmatrix}
  e^{i\frac{2\pi}{l}} &   &   & \\
  & 1 &   & \\
  &   &1  & \\
  & &  & e^{-i\frac{2\pi}{l}}
\end{bmatrix}_{2^{2}\times 2^{2}},
\end{equation*}
\begin{equation*}
\underbrace{\hat{C}_{l}\otimes\cdots \otimes
\hat{C}_{l}}\limits_{l-1}=\begin{bmatrix}
  e^{i\frac{(l-1)}{l}\pi} &   &   & \\
  & * &   & \\
  &   &\cdots  & \\
  & &  & e^{-i\frac{(l-1)}{l}\pi}
\end{bmatrix}_{2^{l-1}\times 2^{l-1}}.
\end{equation*}
Here we only explicitly list the up-left and bottom-right entries
because only these two entries are useful in the following
discussions. The other entries in diagonal are simply represented as
symbol $*$. Note that
\begin{equation*}
\underbrace{\hat{I}\otimes\cdots \otimes
\hat{I}}\limits_{n-l}=\begin{bmatrix}
  1 &   &   & \\
  & 1 &   & \\
  &   &\cdots  & \\
  & &  & 1
\end{bmatrix}_{2^{n-l}\times 2^{n-l}},
\end{equation*}
thus,
\begin{equation*}
\underbrace{\hat{I}\otimes\cdots \otimes
\hat{I}}\limits_{n-l}\otimes\underbrace{\hat{C}_{l}\otimes\cdots
\otimes \hat{C}_{l}}\limits_{l-1}=\begin{bmatrix}
  e^{i\frac{(l-1)}{l}\pi} &   &   & \\
  & * &   & \\
  &   &\cdots  & \\
  & &  & e^{-i\frac{(l-1)}{l}\pi}
\end{bmatrix}_{2^{n-1}\times 2^{n-1}}.
\end{equation*}
Suppose the $n$-th agent chooses arbitrary parameters $(\theta,
\phi)$ in his strategy $(\theta, \phi, a_{n}, t_{n}, z_{n})$, let
\begin{equation*}
  \hat{\omega}(\theta, \phi)=\begin{bmatrix}
  e^{i\phi}\cos(\theta/2) & i\sin(\theta/2)\\
  i\sin(\theta/2) & e^{-i\phi}\cos(\theta/2)
\end{bmatrix},
\end{equation*}
then,
\begin{align*}
\underbrace{\hat{I}\otimes\cdots \otimes
\hat{I}}\limits_{n-l}&\otimes\underbrace{\hat{C}_{l}\otimes\cdots
\otimes \hat{C}_{l}}\limits_{l-1}\otimes\hat{\omega}(\theta,
\phi)\\
&=\begin{bmatrix}
  e^{i[\frac{(l-1)\pi}{l}+\phi]}\cos(\theta/2) &*   &   &   &   &    & \\
  ie^{i\frac{(l-1)\pi}{l}}\sin(\theta/2) &*   &   &   &   &    & \\
   &  & *  & *  &   &    & \\
   &  & *  & *  &   &    & \\
   &  &    &    &\cdots&    & \\
   &  &    &    & & * & ie^{-i\frac{(l-1)\pi}{l}}\sin(\theta/2) \\
   &  &    &    & & * & e^{-i[\frac{(l-1)\pi}{l}+\phi]}\cos(\theta/2)
\end{bmatrix}_{2^{n}\times 2^{n}}.
\end{align*}
Recall that
\begin{equation*}
\overrightarrow{\psi}_{1}=\frac{1}{\sqrt{2}}\begin{bmatrix}
  1\\
  0\\
  \cdots\\
  0\\
  i
\end{bmatrix}_{2^{n}\times 1},
\end{equation*}
thus,
\begin{align*}
\overrightarrow{\psi}_{2}=[\underbrace{\hat{I}\otimes\cdots \otimes
\hat{I}}\limits_{n-l}\otimes\underbrace{\hat{C}_{l}\otimes\cdots
\otimes \hat{C}_{l}}\limits_{l-1}\otimes\hat{\omega}(\theta,
\phi)]\overrightarrow{\psi}_{1}=\frac{1}{\sqrt{2}}\begin{bmatrix}
  e^{i[\frac{(l-1)\pi}{l}+\phi]}\cos(\theta/2)\\
  ie^{i\frac{(l-1)\pi}{l}}\sin(\theta/2)\\
  0\\
  \cdots\\
  0\\
  -e^{-i\frac{(l-1)\pi}{l}}\sin(\theta/2)\\
  ie^{-i[\frac{(l-1)\pi}{l}+\phi]}\cos(\theta/2)
\end{bmatrix}_{2^{n}\times 1},
\end{align*}
\begin{align*}
\overrightarrow{\psi}_{3}=\hat{J}^{+}\overrightarrow{\psi}_{2}&=\frac{1}{2}\begin{bmatrix}
  e^{i[\frac{(l-1)\pi}{l}+\phi]}\cos(\theta/2) + e^{-i[\frac{(l-1)\pi}{l}+\phi]}\cos(\theta/2)\\
  ie^{i\frac{(l-1)\pi}{l}}\sin(\theta/2) + ie^{-i\frac{(l-1)\pi}{l}}\sin(\theta/2)\\
  0\\
  \cdots\\
  0\\
  e^{i\frac{(l-1)\pi}{l}}\sin(\theta/2)-e^{-i\frac{(l-1)\pi}{l}}\sin(\theta/2)\\
  -ie^{i[\frac{(l-1)\pi}{l}+\phi]}\cos(\theta/2) + ie^{-i[\frac{(l-1)\pi}{l}+\phi]}\cos(\theta/2)
\end{bmatrix}_{2^{n}\times 1}\\
&=\quad\begin{bmatrix}
  \cos(\theta/2)\cos(\frac{l-1}{l}\pi+\phi)\\
  i\sin(\theta/2)\cos\frac{l-1}{l}\pi\\
  0\\
  \cdots\\
  0\\
  i\sin(\theta/2)\sin\frac{l-1}{l}\pi \\
  \cos(\theta/2)\sin(\frac{l-1}{l}\pi+\phi)
\end{bmatrix}_{2^{n}\times 1}.
\end{align*}
The probability distribution $\Delta$ is computed from
$\overrightarrow{\psi}_{3}$:
\begin{align}
&P_{C\cdots CC}=\cos^{2}(\theta/2)\cos^{2}(\phi-\frac{\pi}{l})\\
&P_{C\cdots CD}=\sin^{2}(\theta/2)\cos^{2}\frac{\pi}{l}\\
&P_{D\cdots DC}=\sin^{2}(\theta/2)\sin^{2}\frac{\pi}{l}\\
&P_{D\cdots DD}=\cos^{2}(\theta/2)\sin^{2}(\phi-\frac{\pi}{l})
\end{align}
Obviously,
\begin{equation*}
P_{C\cdots CC}+P_{C\cdots CD}+P_{D\cdots DC}+P_{D\cdots DD}=1.
\end{equation*}
Consider the payoff to the $n$-th agent,
\begin{equation}
\$_{n}=\$_{C\cdots CC}P_{C\cdots CC}+\$_{C\cdots CD}P_{C\cdots CD}
  +\$_{D\cdots DC}P_{D\cdots DC}+\$_{D\cdots DD}P_{D\cdots DD}.
\end{equation}
Since $\lambda_{4}$ is satisfied, \emph{i.e.}, $\$_{C\cdots
CC}>\$_{D\cdots DD}$, then the $n$-th agent chooses $\phi=\pi/l$ to
minimize $\sin^{2}(\phi-\frac{\pi}{l})$. As a result, $P_{C\cdots
CC}=\cos^{2}(\theta/2)$.

Since $\lambda_{5}$ is satisfied, \emph{i.e.}, $\$_{C\cdots
CC}>\$_{C\cdots CD}\cos^{2}(\pi/l)+\$_{D\cdots DC}\sin^{2}(\pi/l)$,
then the $n$-th agent prefers $\theta=0$, which leads $\$_{n}$ to
its maximum $\$_{C\cdots CC}$. Therefore, the optimal value of
$(\theta,\phi)$ for the $n$-th agent is
$(0,\pi/l)$.$\quad\quad\quad\quad\quad\quad\quad\quad\quad\square$

Note: The proof of Lemma 1 is similar to the derivation of Eq. (25)
\cite{Flitney2007}.

\newpage
\begin{figure}[!t]
\centering
\includegraphics[height=3.3in,clip,keepaspectratio]{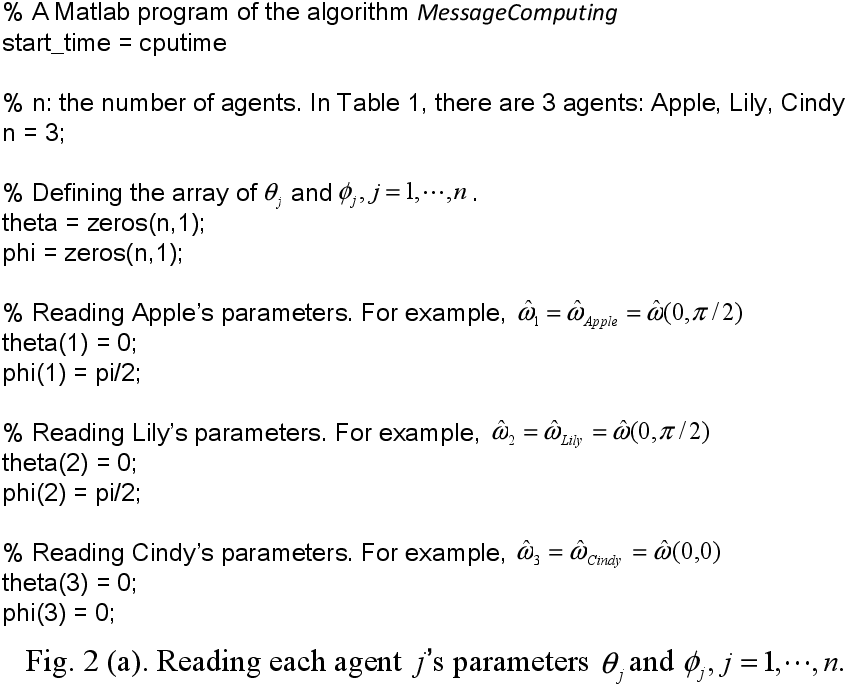}
\end{figure}

\begin{figure}[!t]
\centering
\includegraphics[height=4.9in,clip,keepaspectratio]{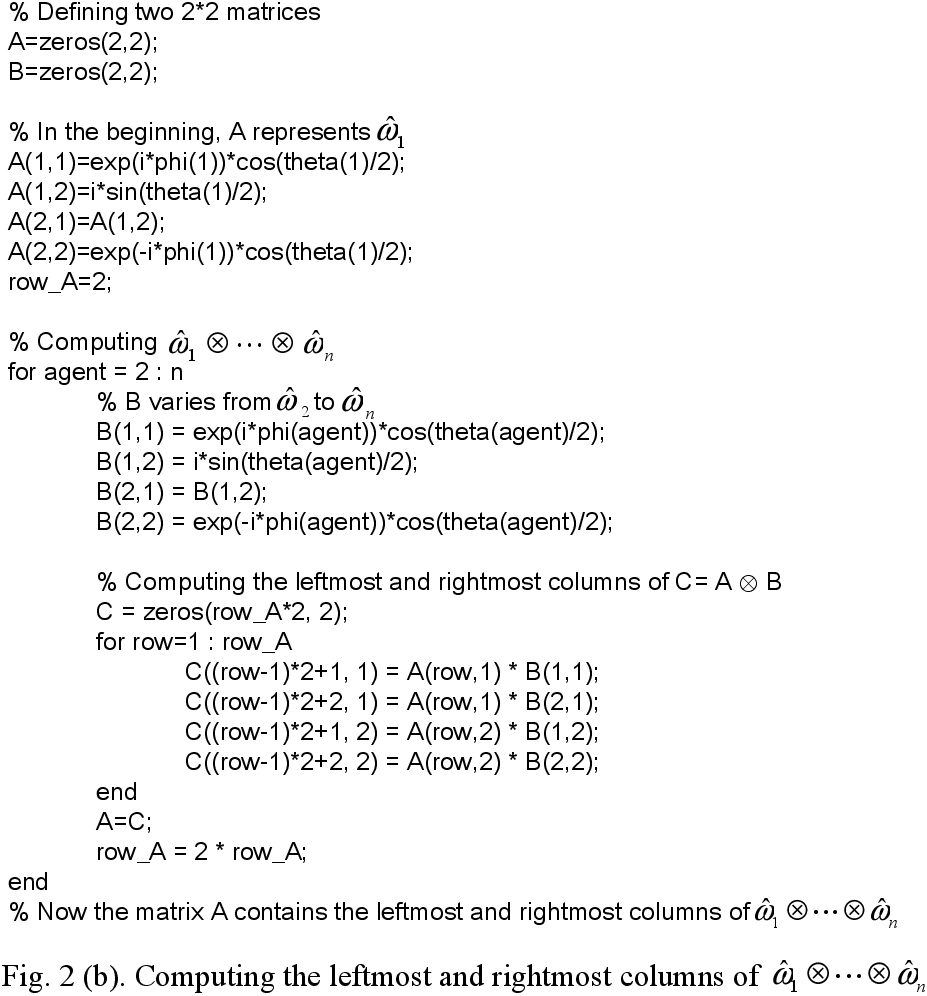}
\end{figure}

\begin{figure}[!t]
\centering
\includegraphics[height=2.4in,clip,keepaspectratio]{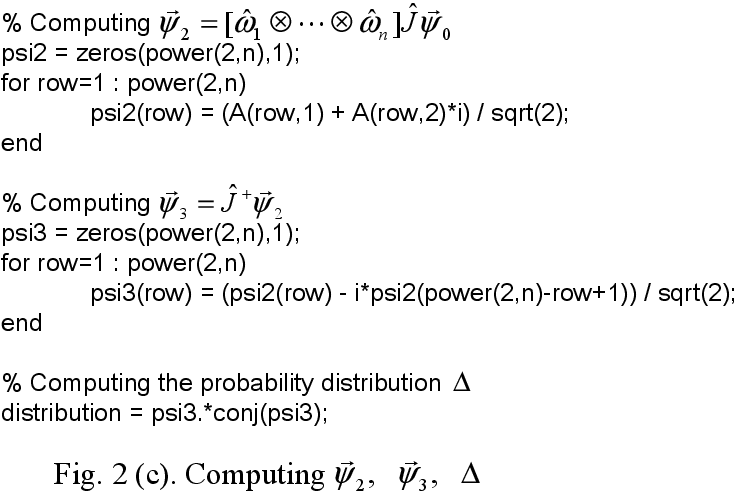}
\end{figure}

\begin{figure}[!t]
\centering
\includegraphics[height=5.7in,clip,keepaspectratio]{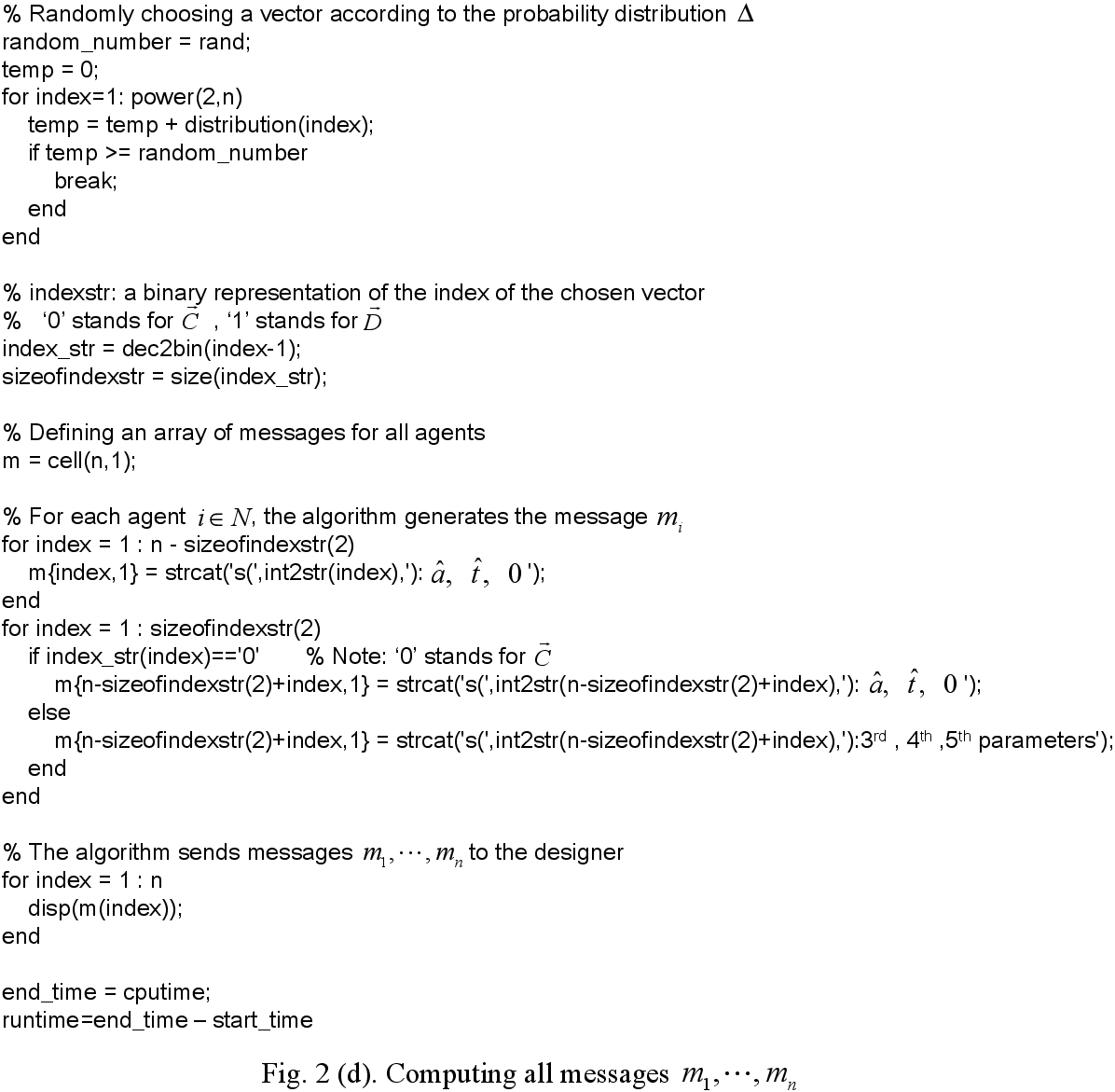}
\end{figure}

\begin{thebibliography}{99}
\bibitem{Maskin1999}
E. Maskin, Nash equilibrium and welfare
optimality, \emph{Rev. Econom. Stud.} \textbf{66} (1999) 23-38.

\bibitem{Serrano2004} R. Serrano, The theory of implementation of
social choice rules, \emph{SIAM Review} \textbf{46} (2004) 377-414.

\bibitem{Telser1980}
L.G. Telser, A theory of self-enforcing agreements. \emph{Journal of
Business} \textbf{53} (1980) 27-44.

\bibitem{Flitney2007}
A.P. Flitney and L.C.L. Hollenberg, Nash equilibria in quantum games
with generalized two-parameter strategies, \emph{Phys. Lett. A}
\textbf{363} (2007) 381-388.
\end{thebibliography}
\end{document}